\documentclass[a4paper,11pt]{article}
\usepackage{pos}

\title{Astrophysical aspects of string compactifications}

\author*[a]{Mario Ramos-Hamud}

\affiliation[a]{DAMTP, University of Cambridge,\\
  Wilberforce Road, Cambridge, CB3 0WA, UK}

\emailAdd{mr895@cam.ac.uk}

\abstract{A generic aspect of low-energy effective field theories (EFTs) coming from string compactifications is the appearance of moduli fields. Among these moduli, the axion and dilaton are present as (pseudo-) Goldstone bosons from the spontaneous breaking of an exact (or approximate) global symmetry. These moduli have a different microscopic coupling to matter but appear kinetically coupled in such a way that their interaction can compete with gravity at low energies and have an important effect in strong gravity environments. In this talk, we will discuss some of the astrophysical implications of a stringy-inspired multi-scalar-tensor theory. In particular, we show the numerical solution of the Tolman-Oppenheimer-Volkov (TOV) system of equations, necessary to probe the existence of a screening mechanism that reduces the Brans-Dicke dilaton coupling to macroscopic matter sources such as a neutron star.}

\FullConference{3rd General Meeting of the COST Action COSMIC WISPers (CA21106) - 3rd Training School of the COST Action COSMIC WISPers (CA21106) (COSMICWISPers2025)\\
9–12 Sept 2025 and 16–19 Sept 2025\\
Sofia, Bulgaria and Annecy, France\\}

\begin{document}
\maketitle

\section{Axio-dilaton scenario}
Motivated by Lovelock's theorem and Occam's razor, the simplest extension of General Relativity (GR) corresponds to adding a single light scalar field minimally coupled to gravity. However, one of the complications of light scalar fields is that they lead to long-range fifth forces, unless a screening mechanism exists to make them consistent with solar system and local tests \cite{Bertotti:2003rm, Will:2014kxa, Berge:2017ovy}. Most of these mechanisms rely on specific features of the scalar potential or zero-derivative matter couplings (see, for example: \cite{Pourhasan:2011sm, Khoury:2003rn}). However, non-derivative terms have been shown to make quantum corrections relevant, possibly invalidating the EFT in the semi-classical regime \cite{Burgess:2009ea, Smith:2025grk} where two-derivative interactions are important.  This requires a very small scalar potential for the scalar fields to compete with gravitational interactions. This sort of light scalars have an origin via spontaneous breaking of a global symmetry \cite{PhysRevLett.29.1698} where they appear as (pseudo)-Goldstone bosons.   It is highly important to notice that a single scalar field is blind to new two-derivative interactions, but when at least two scalar fields are involved, a non-trivial kinetic coupling may appear as a curved target space metric. Inspired by string theory compactifications, where moduli controlling the shape and size of extra dimensions correspond to scalar fields in the four-dimensional low-energy limit, we examine the simplest and minimal case where the EFT only contains one complex modulus, the \textit{axio-dilaton}.  

The most general low-energy effective action for two scalar fields with a curved target space metric  and matter sector is given by
\begin{equation}
	S=  \int d^4x \sqrt{-g} \left\{ \frac{M^2_p}{2} \left[R - (\partial \varphi)^2 - W^2(\varphi)(\partial \mathfrak{a})^2   \right]  - V_{\text{eff}}(\mathfrak{a}) \right\} +S_m[\tilde{g}_{\mu \nu}, \mathfrak{a}, \psi],
\end{equation}
with the Planck mass  $M^2_p =1/ 8 \pi G$, where $G$ is the Newton's constant, $R$ is the 4D Ricci scalar; $\varphi$ is called the dilaton, and $\mathfrak{a}$ the axion which \textit{a priori} is gifted with an internal shift symmetry. Inspired by string theory constructions  \cite{Burgess:2021obw}, we adopt the kinetic coupling of the fields to be $W(\varphi)= e^{- \xi \phi}$. In addition, we denote all the ordinary matter fields with $\psi$ and the dilaton only couples to matter through the Jordan frame metric $\tilde{g}_{\mu \nu}= \exp(2 \mathfrak{g}\varphi)g_{\mu \nu}$. This leads to the equations of motion,
\begin{subequations}
\begin{equation}
    G_{\mu \nu} -  \partial_\mu \varphi \partial_\nu \varphi - W^2 \partial_\mu \mathfrak{a} \partial_\nu \mathfrak{a} + \frac{1}{2} g_{\mu \nu} \left[ (\partial \varphi)^2 +  W^2 (\partial \mathfrak{a})^2+ \frac{2V}{M_p^2}  \right]  = \frac{1}{M_p^2}  T_{\mu \nu}
\label{eq:Eins-ax-screening}
\end{equation}
\begin{equation}
    W^2 \Box a + 2 W W_\varphi \partial^\mu \mathfrak{a} \partial_\mu \varphi- \frac{1}{M_p^2} (V_\mathfrak{a} +U_\mathfrak{a})=0 \quad  \text{and} \quad \Box \varphi -  W W_\varphi (\partial \mathfrak{a})^2+ \frac{\mathfrak{g}T}{M_p ^2} =0,
\end{equation}
\end{subequations}
with $T_{\mu \nu}$, the energy-momentum tensor. We assume spherical symmetry since we are interested in the interaction of the scalar fields with a spherical matter source. Moreover, we consider that the axion shift-symmetry is broken by the macroscopic axion-matter coupling. This leads to the effective axion potential oscillating around its minimum, which can be approximated by
\begin{equation}
	V_\text{eff}(\mathfrak{a}) =  V(\mathfrak{a})+ U(\mathfrak{a}) = \frac{1}{2} \mu_{\text{out}}^2 M^2_p (\mathfrak{a}- \mathfrak{a_+})^2 + \frac{1}{2} \mu_{\text{in}}^2 M^2_p (\mathfrak{a}- \mathfrak{a_+})^2 \varepsilon(r),
\end{equation}
with $\mu_\mathfrak{a}= W(\varphi) m_\mathfrak{a}$, where $m_\mathfrak{a}$ is the axion mass and $\varepsilon(r)$ is the density of the local gravity source the scalar fields are interacting with. The idea is to have different minima inside and outside the source ($\mathfrak{a_-}\neq \mathfrak{a_+}$) to allow the existence of an axion gradient that ultimately can control the screening. This system was previously studied in \cite{Brax:2023qyp} within the \textit{adiabatic approximation} and for compact objects such as the Sun or Earth, using an approximate analytic solution for the axion in the limit when its Compton wavelength is smaller than the radius of the compact object, allowing the axion to transition between minima of the local potential. Instead, here we choose a numerical approach and do not consider any approximation. Moreover, we choose the matter source to be a neutron star.

\section{Numerical solutions}

\begin{figure}
    \centering
    \includegraphics[width=0.8\linewidth]{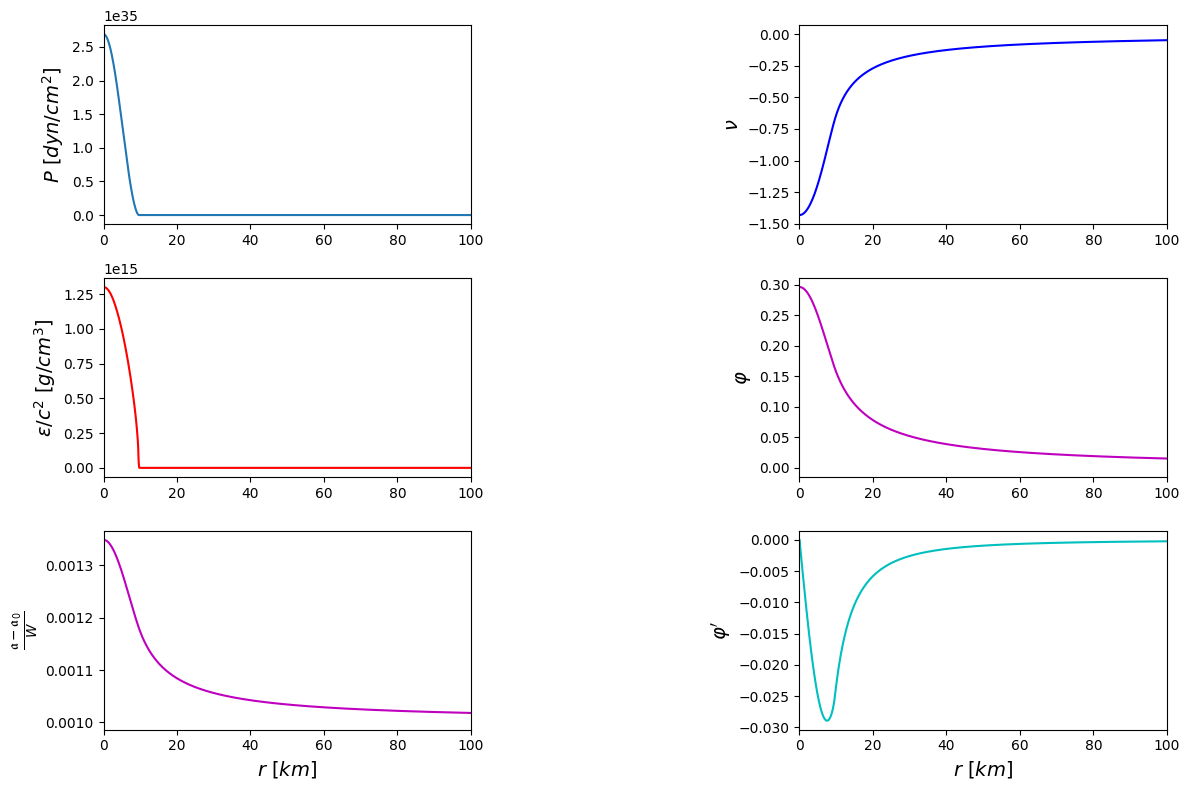}
    \caption{Numerical solution of the TOV-scalar system of equations. Here, we consider $m_\mathfrak{a}= 10^{-15}$eV outside the neutron star.}
    \label{fig:preliminary}
\end{figure}

We numerically solved the TOV-scalar system of equations using a variation of the open code pyTOV-ST  \cite{stergioulas_github} appropriately modified to include the new elements such as the axion, its potential and the kinetic coupling. In addition, we assume a piecewise polytropic equation of state motivated by the study of neutron stars \cite{Douchin:2001sv,Odintsov:2021nqa}, where in the Einstein frame we write the energy density as $\varepsilon$ and the rest-mass density as $\rho$. In this setup \cite{Read:2008iy, Read:2009yp}, the behaviour of the rest-mass density is modelled by a low-density region $\rho< \rho_0$ and a high-energy density region $\rho\gg \rho_0$, for a given $\rho_0$ that connects both regions smoothly. In addition, two (dividing) high densities are considered as a reference, $\rho_1= 10^ {14.7}$ g/cm$^3$ and $\rho_2= 10^ {15.0}$ g/cm$^3$, such that for each region of the piecewise density, $\rho_{i-1} \leq \rho \leq \rho_{i}$, the polytropic equation of state $ P(\rho)= K_i \rho^{\Gamma_i},$ holds, where $K_i$ are constants and $\Gamma_i$ are the adiabatic indices of the type Skyrme-Lyon (SLy) as in \cite{Douchin:2001sv}. The numerical solutions of the TOV-scalar system of equations are shown in Fig. \ref{fig:preliminary}. As expected, the pressure and energy density vanish outside the star, meanwhile, $\nu$, $\varphi$ and $\varphi'$ do it asymptotically. Additionally, the axion field transitions from one minimum to the other, generating a non-vanishing gradient. 

\section{Multi-field screening mechanism}

As suggested in \cite{Smith:2024ayu}, to analyse the existence of screening, we can study the behaviour of the dilaton exterior solution evaluated at the surface of the star, $\varphi_{\text{ext}}'(R)$, sometimes referred to as the dilaton charge. In other words, for $0\leq r<R$ and in the non-relativistic limit, $T\simeq - \rho $, the dilaton's equation can be written as
\begin{equation}
    \varphi' (r) \simeq  \frac{1}{M^2_p r^2} \int_0^r d \hat{r}\; \hat{r}^2  [\mathfrak{g}  \,\rho(\hat{r}) + W W'\mathfrak{a}'^2],
\label{eq:dilaton-derivative}
\end{equation}
and for $W'<0$, the presence of the axion gradient can reduce the charge of the field. Moreover, outside the star, where the source's mass density vanishes, the dilaton's equation becomes simply $(r^2 \varphi'_{\text{ext}})'=0$ (assuming the axion gradient vanishes outside the source) and therefore
\begin{equation}
    \varphi_{\text{ext}}(r)= \varphi_\infty- \frac{L}{r},
\end{equation}
with $\varphi_\infty$ and  $L$ integrations constants. It is important to note that the quantity to test the screening is $(r^2 \varphi'_{\text{ext}})_{r=R}= L$.  If this screening works, the value of $\varphi_{\text{ext}}'(R)$ will be smaller in comparison with the solution inside the star, even without imposing a very restricting constraint on $\mathfrak{g}$. To better understand that, we can evaluate (\ref{eq:dilaton-derivative}) at the surface   using that  $M= 4 \pi \int_0^R d\hat{r} \hat{r}^2 \rho(\hat{r})$, then we get
\begin{equation}
    \varphi'(R) \simeq \frac{2 \mathfrak{g} G M}{R^2}+ \frac{8 \pi G}{R^2} \int_0^R W W' \mathfrak{a}'^2 d\hat{r}.
\end{equation}
In the absence of an axion gradient, the scalar charge can be approximated at the surface by
\begin{equation}
    \varphi'_{NG} \simeq \frac{2 \mathfrak{g}GM}{R^2} \equiv \frac{L_0}{R^2},
 \end{equation}
and we can compute the ratio

\begin{equation}
    \frac{L}{L_0}= \frac{(r^2 \varphi'_{\text{ext}})_{r=R}}{2 \mathfrak{g}GM }= 1 + \frac{4 \pi }{\mathfrak{g}M}\int_0^R W W' \mathfrak{a}'^2 d\hat{r}.
\end{equation}
This ratio is essential since the axion profile at the surface makes the scalar couple to the microscopic sources `as if' its BD coupling were
\begin{equation}
    \mathfrak{g}_{\text{eff}} := \frac{L}{2 GM}=  \mathfrak{g} \frac{L}{L_0} = \mathfrak{g}+ \frac{4 \pi}{M} \int_0^R W W' \mathfrak{a}'^2 d\hat{r}.
    \label{eq:g_eff}
\end{equation}
As expected, this reduces to $\mathfrak{g}$ when the axion gradient vanishes and for a non-trivial axion gradient, it decreases its value by the fact that $W'<0$. Thus, this is the quantity that will enable us to determine whether the screening is effective or not and this is work in progress.

\section{Conclusions}

After having numerically solved the TOV-scalar system of equations for a neutron star with a realistic equation of state, we can proceed to numerically compute the effect of the axion gradient to quantitatively measure its impact on screening the dilaton coupling to matter.  This represents the final phase of this ongoing project and is expected to be completed shortly.

\section*{Acknowledgments}
We would like to thank Philippe Brax, Clifford Burgess and Fernando Quevedo for the ongoing collaboration on this project. This article is based upon work from COST Action COSMIC WISPers CA21106, supported by COST (European Cooperation in Science and Technology).

\bibliographystyle{JHEP}
\bibliography{Literatur}

@article{Burgess:2021obw,
    author = "Burgess, C. P. and Dineen, Danielle and Quevedo, F.",
    title = "{Yoga Dark Energy: natural relaxation and other dark implications of a supersymmetric gravity sector}",
    eprint = "2111.07286",
    archivePrefix = "arXiv",
    primaryClass = "hep-th",
    reportNumber = "CERN-TH-2021-192",
    doi = "10.1088/1475-7516/2022/03/064",
    journal = "JCAP",
    volume = "03",
    number = "03",
    pages = "064",
    year = "2022"
}

@article{Odintsov:2021nqa,
    author = "Odintsov, S. D. and Oikonomou, V. K.",
    title = "{Neutron stars in scalar\textendash{}tensor gravity with quartic order scalar potential}",
    eprint = "2104.01982",
    archivePrefix = "arXiv",
    primaryClass = "gr-qc",
    doi = "10.1016/j.aop.2022.168839",
    journal = "Annals Phys.",
    volume = "440",
    pages = "168839",
    year = "2022"
}

@article{Douchin:2001sv,
    author = "Douchin, F. and Haensel, P.",
    title = "{A unified equation of state of dense matter and neutron star structure}",
    eprint = "astro-ph/0111092",
    archivePrefix = "arXiv",
    doi = "10.1051/0004-6361:20011402",
    journal = "Astron. Astrophys.",
    volume = "380",
    pages = "151",
    year = "2001"
}

@article{Brax:2023qyp,
    author = "Brax, Philippe and Burgess, C. P. and Quevedo, F.",
    title = "{Axio-Chameleons: A Novel String-Friendly Multi-field Screening Mechanism}",
    eprint = "2310.02092",
    archivePrefix = "arXiv",
    primaryClass = "hep-th",
    month = "10",
    year = "2023"
}

@article{Bertotti:2003rm,
    author = "Bertotti, B. and Iess, L. and Tortora, P.",
    title = "{A test of general relativity using radio links with the Cassini spacecraft}",
    doi = "10.1038/nature01997",
    journal = "Nature",
    volume = "425",
    pages = "374--376",
    year = "2003"
}

@article{Smith:2024ayu,
    author = "Smith, Adam and Mylova, Maria and Brax, Philippe and van de Bruck, Carsten and Burgess, C. P. and Davis, Anne-Christine",
    title = "{CMB Implications of Multi-field Axio-dilaton Cosmology}",
    eprint = "2408.10820",
    archivePrefix = "arXiv",
    primaryClass = "hep-th",
    month = "8",
    year = "2024"
}

@article{Read:2009yp,
    author = "Read, Jocelyn S. and Markakis, Charalampos and Shibata, Masaru and Uryu, Koji and Creighton, Jolien D. E. and Friedman, John L.",
    title = "{Measuring the neutron star equation of state with gravitational wave observations}",
    eprint = "0901.3258",
    archivePrefix = "arXiv",
    primaryClass = "gr-qc",
    doi = "10.1103/PhysRevD.79.124033",
    journal = "Phys. Rev. D",
    volume = "79",
    pages = "124033",
    year = "2009"
}

@article{Read:2008iy,
    author = "Read, Jocelyn S. and Lackey, Benjamin D. and Owen, Benjamin J. and Friedman, John L.",
    title = "{Constraints on a phenomenologically parameterized neutron-star equation of state}",
    eprint = "0812.2163",
    archivePrefix = "arXiv",
    primaryClass = "astro-ph",
    doi = "10.1103/PhysRevD.79.124032",
    journal = "Phys. Rev. D",
    volume = "79",
    pages = "124032",
    year = "2009"
}

@article{Will:2014kxa,
    author = "Will, Clifford M.",
    title = "{The Confrontation between General Relativity and Experiment}",
    eprint = "1403.7377",
    archivePrefix = "arXiv",
    primaryClass = "gr-qc",
    doi = "10.12942/lrr-2014-4",
    journal = "Living Rev. Rel.",
    volume = "17",
    pages = "4",
    year = "2014"
}

@article{Berge:2017ovy,
    author = "Berg{\'e}, Joel and Brax, Philippe and M{\'e}tris, Gilles and Pernot-Borr{\`a}s, Martin and Touboul, Pierre and Uzan, Jean-Philippe",
    title = "{MICROSCOPE Mission: First Constraints on the Violation of the Weak Equivalence Principle by a Light Scalar Dilaton}",
    eprint = "1712.00483",
    archivePrefix = "arXiv",
    primaryClass = "gr-qc",
    doi = "10.1103/PhysRevLett.120.141101",
    journal = "Phys. Rev. Lett.",
    volume = "120",
    number = "14",
    pages = "141101",
    year = "2018"
}

@article{Pourhasan:2011sm,
    author = "Pourhasan, R. and Afshordi, N. and Mann, R. B. and Davis, A. C.",
    title = "{Chameleon Gravity, Electrostatics, and Kinematics in the Outer Galaxy}",
    eprint = "1109.0538",
    archivePrefix = "arXiv",
    primaryClass = "astro-ph.CO",
    doi = "10.1088/1475-7516/2011/12/005",
    journal = "JCAP",
    volume = "12",
    pages = "005",
    year = "2011"
}

@article{Khoury:2003rn,
    author = "Khoury, Justin and Weltman, Amanda",
    title = "{Chameleon cosmology}",
    eprint = "astro-ph/0309411",
    archivePrefix = "arXiv",
    doi = "10.1103/PhysRevD.69.044026",
    journal = "Phys. Rev. D",
    volume = "69",
    pages = "044026",
    year = "2004"
}

@article{PhysRevLett.29.1698,
  title = {Approximate Symmetries and Pseudo-Goldstone Bosons},
  author = {Weinberg, Steven},
  journal = {Phys. Rev. Lett.},
  volume = {29},
  issue = {25},
  pages = {1698--1701},
  numpages = {0},
  year = {1972},
  month = {Dec},
  publisher = {American Physical Society},
  doi = {10.1103/PhysRevLett.29.1698},
  url = {https://link.aps.org/doi/10.1103/PhysRevLett.29.1698}
}

@article{Burgess:2009ea,
    author = "Burgess, C. P. and Lee, Hyun Min and Trott, Michael",
    title = "{Power-counting and the Validity of the Classical Approximation During Inflation}",
    eprint = "0902.4465",
    archivePrefix = "arXiv",
    primaryClass = "hep-ph",
    reportNumber = "PI-PARTPHYS-121",
    doi = "10.1088/1126-6708/2009/09/103",
    journal = "JHEP",
    volume = "09",
    pages = "103",
    year = "2009"
}

@article{Smith:2025grk,
    author = "Smith, Adam and Brax, Philippe and van de Bruck, Carsten and Burgess, C. P. and Davis, Anne-Christine",
    title = "{Screened axio-dilaton cosmology: novel forms of early dark energy}",
    eprint = "2505.05450",
    archivePrefix = "arXiv",
    primaryClass = "hep-th",
    doi = "10.1140/epjc/s10052-025-14735-4",
    journal = "Eur. Phys. J. C",
    volume = "85",
    number = "9",
    pages = "1062",
    year = "2025"
}

@misc{stergioulas_github,
  author       = {Nikolaos Stergioulas},
  howpublished = {\url{https://github.com/niksterg}},
  note         = {Accessed: 2025-12-21}
}

\end{document}